\documentclass{emulateapj}
\usepackage{times, graphicx}

\shorttitle{The Source of Three-minute Oscillations in Coronal Fans}
\shortauthors{D.B. Jess et al.}

\begin{document}


\title{The Source of Three-minute Magneto-acoustic Oscillations in Coronal Fans}


\author{D. B. Jess$^{1}$, I. De Moortel$^{2}$, M. Mathioudakis$^{1}$, D. J. Christian$^{3}$, \\ 
K. P. Reardon$^{4,1}$, P. H. Keys$^{1}$, F. P. Keenan$^{1}$}
\affil{$^{1}$Astrophysics Research Centre, School of Mathematics and Physics, 
Queen's University Belfast, Belfast, BT7~1NN, Northern Ireland, U.K.}
\affil{$^{2}$School of Mathematics and Statistics, University of St Andrews, St Andrews, 
Scotland, KY16 9SS, UK}
\affil{$^{3}$Department of Physics and Astronomy, California State University Northridge, 
Northridge, CA 91330, U.S.A.}
\affil{$^{4}$Osservatorio Astrofisico di Arcetri, 50125 Firenze, Italy}
\email{d.jess@qub.ac.uk}




\begin{abstract}
We use images of high spatial, spectral and temporal resolution, obtained 
using both ground- and space-based instrumentation, to 
investigate the coupling between wave phenomena observed at numerous 
heights in the solar atmosphere. Analysis of $4170${\,}{\AA} 
continuum images reveals small-scale umbral intensity enhancements, 
with diameters $\sim$$0{\,}.{\!\!}{\arcsec}6$, lasting in excess of $30$~minutes. 
Intensity oscillations of $\approx$$3$~minutes are observed to encompass 
these photospheric structures, with power at least three 
orders-of-magnitude higher than the surrounding umbra. 
Simultaneous chromospheric velocity and intensity time series reveal 
an $87\pm8\degr$ out-of-phase behavior, implying the presence of 
standing modes created as a result of partial wave reflection at the transition region 
boundary. We find a maximum wave guide inclination angle 
of $\approx$$40\degr$ between photospheric and chromospheric heights, 
combined with a radial expansion factor of $<$$76$\%. An average blue-shifted 
Doppler velocity of $\approx$$1.5$~km{\,}s$^{-1}$, in addition to a time lag between 
photospheric and chromospheric oscillatory phenomena, 
confirms the presence of upwardly-propagating 
slow-mode waves in the lower solar atmosphere. 
Propagating oscillations in EUV intensity are detected in simultaneous 
coronal fan structures, with a periodicity of $172\pm17$~s and a 
propagation velocity of $45\pm7$~km{\,}s$^{-1}$. 
Numerical simulations reveal that the damping of the magneto-acoustic wave 
trains is dominated by thermal conduction. The coronal 
fans are seen to anchor into the photosphere in locations where 
large-amplitude umbral dot oscillations manifest. Derived 
kinetic temperature and emission measure time-series display 
prominent out-of-phase characteristics, and when combined with the 
previously established sub-sonic wave speeds, we conclude that the observed 
EUV waves are the coronal counterparts of the upwardly-propagating 
magneto-acoustic slow-modes detected in the lower solar atmosphere. 
Thus, for the first time, we reveal how the propagation of $3$~minute 
magneto-acoustic waves in solar coronal structures is a direct result of amplitude 
enhancements occurring in photospheric umbral dots. 
\end{abstract}



\section{Introduction}
Early white-light eclipse photographs of the Sun revealed elongated, 
faint columns of enhanced density stretching far out into the corona 
\citep{van50, Sai65}. These structures, now commonly 
referred to as coronal plumes, can be viewed over a wide range of 
wavelengths, in particular the extreme ultraviolet 
\citep[EUV;][]{Boh75, Ahm77}. Plumes are just one type of 
EUV feature that are seen to exist in the solar corona. Other 
examples include coronal loop and fan structures, which are 
observed to outline the coronal magnetic-field topology, and 
demonstrate a wide range of oscillatory behaviour 
\citep[{\rmfamily e.g.,}][]{Asc99, Asc02, Nak99, Jes08c, Ofm08, Ver08, Van08, Bal11}. 

One of the first studies which uncovered propagating wave 
phenomena in coronal structures was that by \citet{Def98}. These 
authors utilised the Extreme-ultraviolet Imaging Telescope on 
board the Solar and Heliospheric Observatory 
spacecraft to identify quasi-periodic perturbations in 
the brightness of $171$~{\AA} images. \citet{DeM00} 
undertook a similar study using the higher spatial resolution 
Transition Region and Coronal Explorer \citep[TRACE;][]{Han99}, 
and concluded that these oscillations were signatures of 
slow magneto-acoustic waves, which propagate upwards 
along the coronal waveguides with velocities of 
$70$--$165$~km{\,}s$^{-1}$ and periods in the range 
$180$--$420$~s. Energy estimates for 
these motions exhibit an incredibly wide range of values, 
typically $10^{2}$--$10^{5}$~ergs{\,}cm$^{-2}${\,}s$^{-1}$ 
\citep{Def98, DeM00}.

Since magnetic fields play an essential 
role in plume/fan/coronal-loop formation and structuring, they are often 
modelled using magnetohydrodynamic (MHD) equations 
\citep{Del97}. Utilising non-linear, two-dimensional 
MHD simulations, \citet{Ofm99} were 
able to replicate previous observational results, and concluded that 
outward-propagating slow magneto-acoustic waves 
may be able to contribute significantly to the heating of 
the lower corona through compressive dissipation. 
Furthermore, theoretical modelling has suggested that 
the propagation characteristics of a magneto-acoustic 
intensity perturbation depends on a number of factors, 
including the dissipation of the wave energy \citep{Kli04, DeM12}. 
Numerical simulations indicate that thermal conduction 
may be the dominant damping mechanism behind the 
dissipation of magneto-acoustic wave energy in the 
solar corona \citep{Ofm02, DeM03, DeM04, Men04}.

\begin{figure*}
\includegraphics[width=\textwidth]{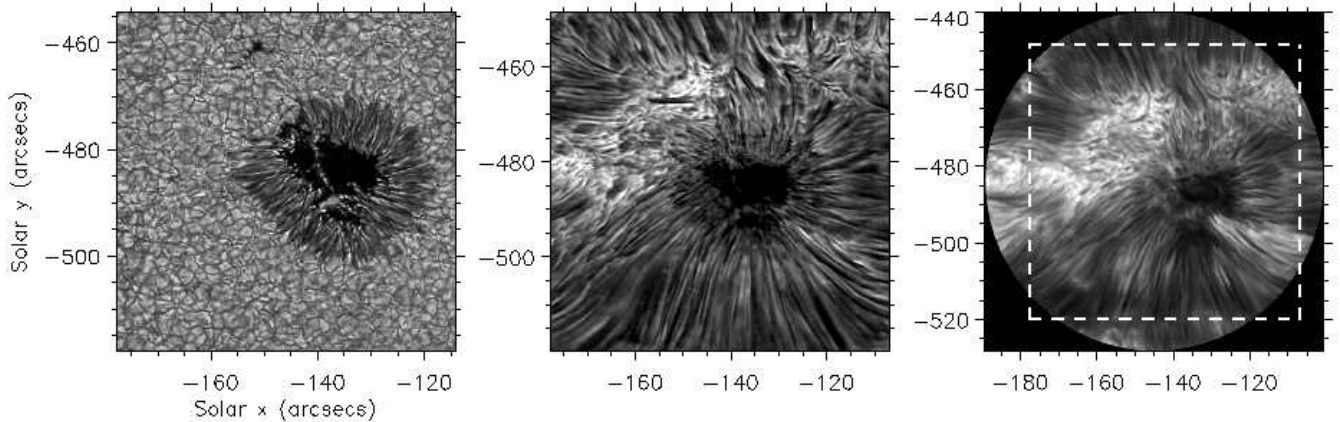}
\caption{Simultaneous images of the lower solar atmosphere, 
obtained through $4170${\,}{\AA} continuum (left), H$\alpha$ core (middle), 
and Ca~{\sc{ii}}~$8542${\,}{\AA} core (right) filters at $13$:$32$~UT on $2011$ 
July $13$. The Ca~{\sc{ii}}~$8542${\,}{\AA} image displays Doppler-compensated intensities, while 
the dashed white box outlines the ROSA/HARDcam field-of-view. 
Axes are in heliocentric arcseconds, where $1${\arcsec} $\approx$ $725$~km on the 
solar surface.}
\label{fig1}
\end{figure*}

Through examination of coronal structures in the close 
proximity of active regions, \citet{Flu01} 
and \citet{Mar06} were able to 
reveal how structures situated above sunspot regions 
displayed intensity oscillations with a period of the 
order of three minutes, while oscillations in ``non-sunspot'' 
structures demonstrated much longer periodicities 
\citep{DeM02}. The 
authors concluded that the most likely explanation 
for the observed longitudinal waves revolves around 
a driver directly exciting the magnetic footpoints. This 
scenario requires the magneto-acoustic wave trains 
to be able to propagate from the lower solar atmosphere, 
through the transition region, and into the corona. 
Utilising numerical simulations of the Sun's lower 
atmosphere, \citet{Kho06} have revealed how 
longitudinal oscillations, generated with a 
periodicity of $\sim$$3$~minutes in sunspot umbrae, can 
readily propagate upwards from the photosphere and 
into the chromosphere. 

Previously, three minute umbral oscillations have been 
notoriously difficult to detect at photospheric heights. 
\citet{Bal87} were unable to detect any photospheric 
signatures using the Locarno solar station at the 
G{\"o}ttingen Observatory, while \citet{Bru85} suggested that 
they may get lost in the noise, as a result of their 
very low amplitude. Indeed, \citet{Nag07} 
utilised Hinode/SOT image sequences to show how 
oscillatory power, at all frequencies, is significantly 
reduced in sunspot umbrae. More recently, 
\citet{Kob08, Kob11} have not only detected 
photospheric three minute oscillations, but the 
authors also claim that the 
location of maximum chromospheric 
power also corresponds to a co-spatial decrease in power 
of the photospheric oscillations. However, 
the spatial resolution obtained by \citet{Kob08, Kob11} 
was on the order of 1{\arcsec}, so precise diagnostics 
of the exact umbral structures displaying three minute 
periodicities was impossible.

In this paper, we utilise ground- and space-based 
instrumentation, with high spatial, 
temporal and spectral resolution, to investigate the 
origin of $3$~minute magneto-acoustic waves 
observed in EUV images of coronal fan structures. 
We employ a multi-wavelength approach  
to study the photospheric counterpart of these 
coronal phenomena, and analyse the resulting wave 
propagation characteristics from the photosphere, through 
the chromosphere, and out into the corona. 

\newpage
\section{Observations \& Data Preparation}
\subsection{Ground-based Data}

The observational data presented here are part of a sequence obtained during 
$13$:$32$ -- $14$:$04$~UT on $2011$ July $13$, with the Dunn Solar Telescope (DST) at Sacramento 
Peak, New Mexico. We employed the Rapid Oscillations in the Solar Atmosphere 
\citep[ROSA;][]{Jes10b} camera system 
to image a $69${\arcsec} $\times$ $69${\arcsec} region encompassing 
active region NOAA~$11250$, positioned at heliocentric co-ordinates 
($-146${\arcsec}, $-486${\arcsec}), or S$26.3$E$10.1$ in the conventional 
north-south-east-west co-ordinate system.   
A spatial sampling of $0{\,}.{\!\!}{\arcsec}069$ per pixel was used for the ROSA cameras, to 
match the telescope's diffraction-limited resolution in the blue continuum to that of 
the CCD. This results in images obtained at longer wavelengths being slightly 
oversampled. However, this was deemed desirable to keep the dimensions of the 
field-of-view the same for all ROSA cameras. 

A recent addition to the DST's imaging capabilities is a new high quantum 
efficiency device, the Hydrogen-Alpha Rapid Dynamics camera 
(HARDcam). 
This camera, an iXon X3 DU-887-BV\footnote{Full specifications available 
at http://www.andor.com} 
model manufactured by Andor Technology, 
consists of a back-illuminated, $512$$\times$$512$ pixel$^{2}$ 
electron-multiplying CCD, with a quantum efficiency exceeding 95{\%} 
at $6500$~{\AA}. As a result, this camera is best suited 
to imaging in the red portion of the optical spectrum. 
The triggering and readout architectures are identical to the first-generation 
ROSA instrumentation, allowing HARDcam to be seamlessly integrated with the 
existing setup. To optimise the exploitation of its high quantum efficiency, 
HARDcam was employed behind a $0.25$~{\AA} 
H$\alpha$ core filter, incorporating a spatial 
sampling of $0{\,}.{\!\!}{\arcsec}138$ per pixel, providing a 
field-of-view size ($71${\arcsec} $\times$ $71${\arcsec}) 
comparable to existing ROSA image sequences.

\begin{figure*}
\begin{center}
\includegraphics[angle=0,width=12cm]{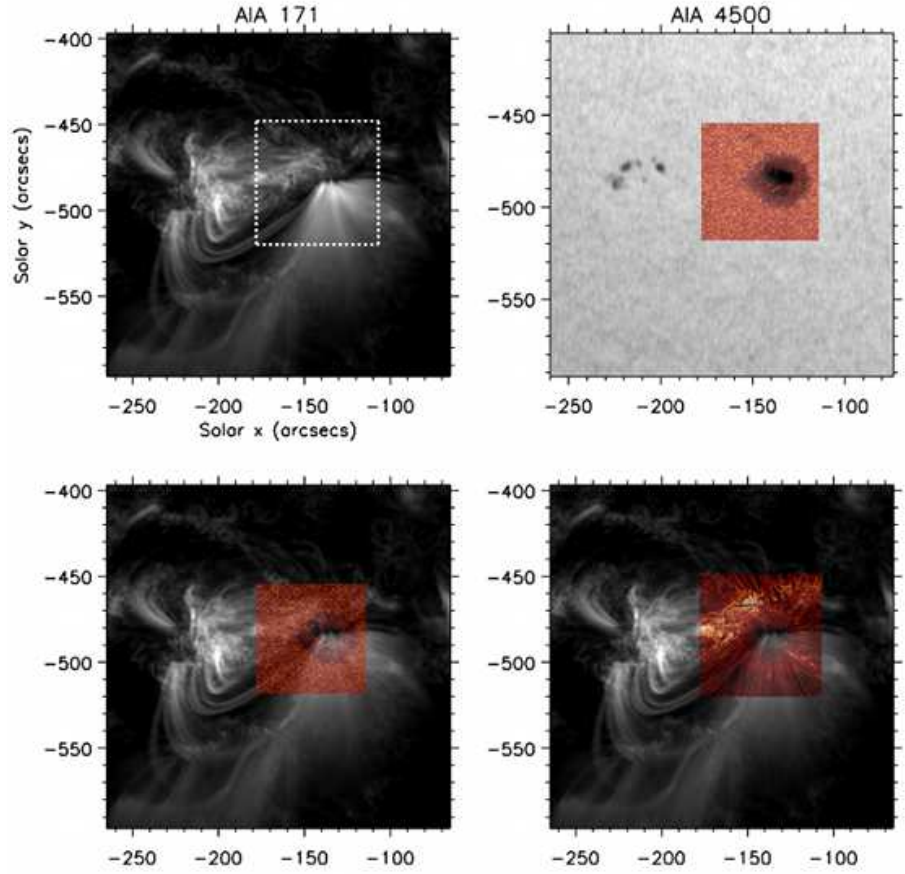}
\caption{{\it{Top Left:}} An AIA $171${\,}{\AA} image, taken 
at $13$:$32$~UT, including a dashed white box to indicate the 
ROSA/HARDcam field-of-view. {\it{Top Right:}} An AIA 
$4500${\,}{\AA} context image, interlaced with a ROSA $4170${\,}{\AA} 
continuum snapshot, following sub-pixel co-alignment. 
{\it{Bottom Panels:}} Excellent co-alignment between 
lower and upper solar atmosphere datasets allow for precise spatial diagnostics, as 
demonstrated by the interlacing of $4170${\,}{\AA} continuum (left) and 
H$\alpha$ core (right) images, with the coronal emission seen in the 
AIA $171${\,}{\AA} bandpass.}
\label{fig2}
\end{center}
\end{figure*}

\begin{table}
\caption{ROSA/HARDcam/IBIS filter and cadence overview. \label{table1}}
\begin{tabular}{lrrr}
~&~&~&~ \\
Filter Used			& Exposure Time 	& Frames per 		& Reconstructed \\
					& (ms)			& Second			& Cadence (s) \\
~&~&~&~ \\
\tableline
Continuum ($4170$~\AA) 	& $7$	 	& $30.3$ 		& $1.06$        \\
H$\alpha$ core			& $35$ 		& $27.9$ 		& $1.26$        \\
Ca~{\sc{ii}} ($8542.12$~\AA)	& $250$		& $^{a}$$3.0$	& $^{b}$$43.4$ \\
~&~&~&~ \\
\end{tabular}
\footnotesize \\
$^{a}$: Average frames per second including Fabry-Perot tuning time \\
$^{b}$: Cadence of a complete Ca~{\sc{ii}} profile scan \\
\end{table}

In addition to ROSA and HARDcam observations, 
the Interferometric BIdimensional Spectrometer \citep[IBIS;][]{Cav06} was used 
to simultaneously sample the Ca~{\sc{ii}} absorption profile at $8542.12$~{\AA}. 
IBIS employed a spatial sampling of $0{\,}.{\!\!}{\arcsec}097$ per pixel, which allowed 
ROSA's near square field-of-view to be contained within the circular aperture 
provided by IBIS. Thirteen discreet wavelength steps, with ten exposures per 
step to assist image reconstruction, were used to provide a complete scan 
cadence of $43.4$~s. A whitelight camera, synchronised with the IBIS 
feed, was utilised to assist the processing of narrowband 
images. Full details of the observations presented here, including filters and 
exposure times used, can be found in Table~\ref{table1}, while sample 
images can be viewed in Figure~\ref{fig1}.

During the observations, high-order adaptive optics \citep{Rim04} 
were used to correct wavefront deformations in real-time. The acquired images were 
further improved through speckle reconstruction algorithms \citep{Wog08}, 
utilizing $32 \rightarrow 1$ restorations for the 
G-band and 4170{\,}{\AA} continuum images. 
The remaining HARDcam and IBIS images were processed with 
$35 \rightarrow 1$ and $10 \rightarrow 1$ restorations, respectively. 
Post-reconstruction cadences are displayed in the fourth 
column of Table~\ref{table1}. A full image-reconstructed IBIS 
scan through the Ca~{\sc{ii}} absorption line includes a blueshift 
correction, required due to the use of classical etalon mountings 
\citep{Cau08}. Atmospheric seeing conditions remained excellent 
throughout the time series. However, to insure accurate co-alignment in all 
bandpasses, broadband time series were Fourier co-registered and de-stretched using a 
$40 \times 40$ grid, equating to a $\approx$$1{\,}.{\!\!}{\arcsec}7$ separation between spatial samples 
\citep{Jes07,Jes08a}. Narrowband images, including those from IBIS, were corrected 
by applying destretching vectors established from simultaneous broadband reference 
images \citep{Rea08, Jes10a, Rea11}. 

\begin{figure*}
\includegraphics[width=\textwidth]{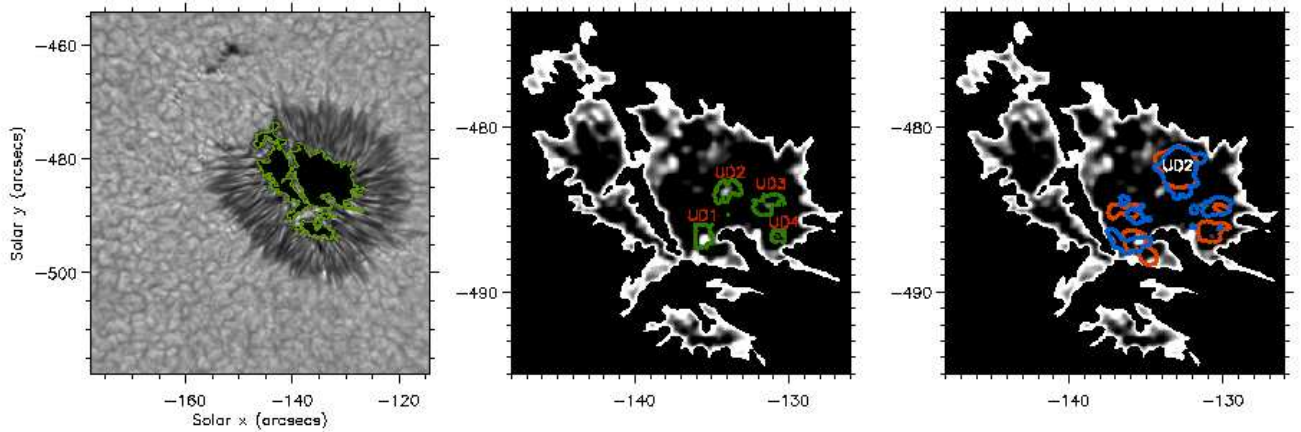}
\caption{{\it{Left:}} A time-averaged $4170${\,}{\AA} continuum 
image, where a solid green contour defines the outline of the 
sunspot umbra. {\it{Middle and Right:}} A time-averaged close-up of the 
sunspot umbra, following isolation from the surrounding 
penumbral and granulation structures. Locations of the umbral dots analysed 
here are indicated by red ``UDx'' markings, where the ``x'' refers to a number. 
Green, red, and blue contours indicate the locations where photospheric 
intensities, chromospheric H$\alpha$/Ca~{\sc{ii}} line core 
intensities, and Doppler velocities, respectively, display oscillatory power 
exceeding $1000$ times the background 
umbral value, for a periodicity of approximately $3$~minutes.}
\label{fig3}
\end{figure*}

\subsection{Space-based Data}
The Atmospheric Imaging Assembly \citep[AIA;][]{Lem11} onboard 
the Solar Dynamics Observatory \citep[SDO;][]{Pes11} 
was utilised to provide simultaneous EUV images of active 
region NOAA~$11250$. This instrument images the 
entire solar disk in 10 different channels, incorporating a two-pixel spatial 
resolution of $1{\,}.{\!\!}{\arcsec}2$, and a cadence of $12$~s. Here, 
we selected $7$~EUV datasets spanning $13$:$30$ -- $14$:$05$~UT on $2011$ July $13$, 
consisting of $175$ images in each of the $94${\,}{\AA}, $131${\,}{\AA}, 
$171${\,}{\AA}, $193${\,}{\AA}, $211${\,}{\AA}, $304${\,}{\AA}, and $335${\,}{\AA} channels. 
In addition, one 
contextual $4500${\,}{\AA} continuum image, acquired at $14$:$00$~UT, was 
obtained for the purposes of co-aligning AIA datasets with images of the 
lower solar atmosphere. 

The EUV bandpasses 
were specifically chosen to cover a wide range of transition region and 
coronal temperatures, spanning approximately $50{\,}000$~K -- $7$~MK, 
under non-flaring conditions. 
Transition region imaging is covered by the He~{\sc{ii}}-dominated 
$304${\,}{\AA} bandpass, with a typical formation temperature of 
$\sim$$50{\,}000$~K. The selected coronal channels, 
consisting of the $94${\,}{\AA}, $131${\,}{\AA}, $171${\,}{\AA}, $193${\,}{\AA}, 
$211${\,}{\AA}, and $335${\,}{\AA} 
bandpasses, demonstrate typical effective temperatures of approximately 
$7.0$~MK, $0.4$~MK, $0.7$~MK, $1.6$~MK, $2.0$~MK, and $2.8$~MK respectively 
\citep{Odw10, Bro11}. Thus, the AIA datasets, in conjunction with our 
ground-based observations of the lower solar atmosphere, provide us with 
the ideal opportunity to investigate the coupling of lower atmospheric 
(photosphere and chromosphere) phenomena with their multi-million 
degree coronal counterparts at unprecedented spatial and temporal 
resolutions.

The AIA data were processed using the standard {\tt{aia\_prep}} 
routine, and include the removal of energetic particle hits, in addition to the 
co-registration of images from different wavelengths on to a common 
plate scale. Subsequently, $200${\arcsec} $\times$ $200${\arcsec} 
sub-fields were extracted from the processed data, 
with a central pointing close to that of the ground-based image 
sequences. A sub-field image, including an outline of the 
field-of-view obtained with ground-based observations, is 
shown in the upper-left panel of Figure~\ref{fig2}. Using the AIA $4500${\,}{\AA} 
context image to define 
absolute solar co-ordinates, our ground-based observations were 
subjected to cross-correlation techniques to provide sub-pixel 
co-alignment accuracy between the imaging sequences. 
To do this, the plate scales of our ground-based observations 
were first degraded to match that of the AIA image\footnote{Data 
analysis was performed on full-resolution ({\rmfamily i.e.} non-degraded) 
image sequences}. Next, 
squared mean absolute deviations were calculated between the 
datasets, with the ground-based images subsequently 
shifted to best align with the AIA reference image (see the lower panels 
of Figure~\ref{fig2} for interlaced examples).
Following co-alignment, the maximum x- and y-displacements 
are both less than one tenth of an AIA pixel, or $0{\,}.{\!\!}{\arcsec}06$ 
($\approx$$45$~km).

\section{Data Analysis and Discussion}
Using a combination of ground- and space-based datasets, we were 
able to analyse image sequences from the photosphere through 
to the corona. Due to the wide range of atmospheric heights 
covered, this section is divided into 
sub-sections in which we discuss different regions of the 
solar atmosphere.

\subsection{Photosphere}
\label{photosphere}
To investigate oscillatory phenomena occurring in the sunspot 
umbra, we first masked out all other areas of the field-of-view. This 
was deemed desirable so that intrinsically faint umbral structures would 
become much more apparent in scaled intensity images. To isolate 
the umbra, we first had to create 
a binary map, whereby umbral pixels were given a value of `$1$', and 
non-umbral pixels assigned a value of `$0$'. This form of binary 
map was created by first averaging the $4170${\,}{\AA} continuum 
images over the entire $32$~minute duration of the dataset. Next, 
the umbral pixels were defined as those with an intensity below 
$45$\% of the median granulation intensity, and subsequently 
assigned a value of `$1$'. This accurately defined the perimeter of the 
dark umbra. However, some small-scale structures, 
which existed inside the umbra ({\rmfamily e.g.,} umbral dots; UDs), exhibited 
higher relative intensity values. Therefore, to include all structures which existed 
inside the umbral boundary, all pixels which lay inside the outer perimeter 
were assigned a value of `$1$', while all other pixels were given a value of `$0$'. 
The time-averaged $4170${\,}{\AA} continuum image, including 
the outline of the umbral binary map, is shown in the left panel of 
Figure~\ref{fig3}. 

\begin{figure}
\includegraphics[width=\columnwidth]{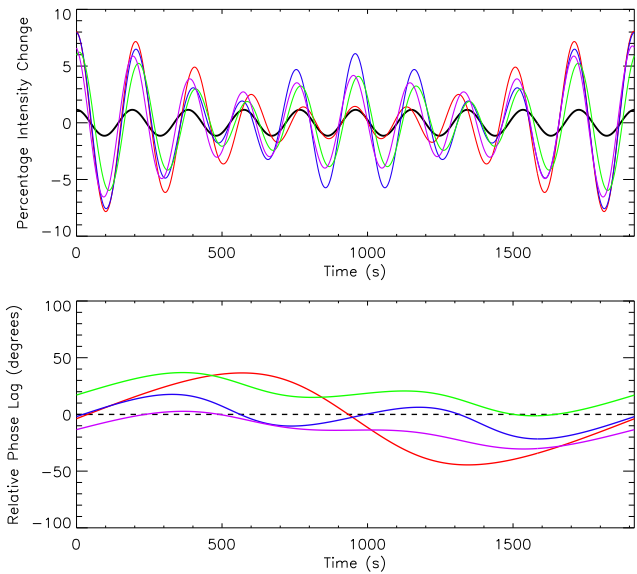}
\caption{{\it{Upper:}} Intensity time series of the four umbral 
dots labelled UD1 (red line), UD2 (blue line), UD3 (green line), 
and UD4 (pink line) in Figure~\ref{fig3}, following temporal 
filtering using a $2$ -- $4$~minute 
bandpass filter. The black line displays the intensity of 
the umbral region away from UD structures, devoid of long-lived or transient 
brightenings, and scaled (by a factor of $5$) for clarity. {\it{Lower:}} The relative 
phase angles between the $3$~minute umbral dot and 
background umbral oscillations, plotted as a function of time. 
The red, blue, green, and pink lines represent umbral dots 
UD1, UD2, UD3, and UD4, respectively, while a black dashed 
line highlights a phase angle of $0$$\degr$.}
\label{fig4}
\end{figure}

Examination of the time-averaged umbral intensity 
map (middle panel of Figure~\ref{fig3}) revealed a collection of 
near-circular brightenings. These brightenings are consistent with 
the signatures of UDs, whereby they typically exhibit diameters 
between $0{\,}.{\!\!}{\arcsec}2$ -- $0{\,}.{\!\!}{\arcsec}8$, and intensities $1.3$ -- $1.5$ 
times brighter than the background umbral value 
\citep{Den98, Tri02, Sob05}. Due to their sharp appearance in a 
time-averaged intensity image, it hints at the persistent 
nature of such structures in the same spatial location, either 
through long lifetimes 
\citep[{\rmfamily e.g.,} $30$~minutes or more;][]{Sob97a, Sob97b}, or by the 
reappearance of new UDs in the same spatial location. 
Analysing simultaneous G-band 
and continuum filtergrams, \citet{Rim08} found that 
UDs, when observed in the G-band, displayed dark 
fine-structuring towards their core. Contrarily, this author found 
that these dark features 
are not as readily visible in the continuum, and concluded this 
may be due to the reduced resolution of the longer wavelength 
and longer exposure images. However, our continuum 
images were obtained at a shorter wavelength than the G-band 
($4170${\,}{\AA} versus $4305${\,}{\AA}), in addition to being 
captured with a much shorter exposure time ($7$~ms versus $20$~ms). 
Thus, we cannot attribute a lack of dark fine-scale structuring 
in the UDs observed in the $4170${\,}{\AA} continuum to either 
reduced spatial resolution, or exposure-time related 
smearing of the images. \citet{Jes12} recently determined the 
contribution functions of the G-band and $4170${\,}{\AA} 
continuum filter bandpases used at the DST, and concluded that 
the blue continuum is formed at a height of $\sim$$25$~km, while the 
G-band is formed approximately $100$~km higher. This atmospheric height 
separation between the two bandpasses, in addition to the appearance 
of diatomic CH molecules in the G-band images, may contribute to the 
presence of dark cores observed in G-band UDs.

Since we are concerned with the 
larger macro-scale fluctuations in umbral intensity, rather than the 
very small-scale structuring ({\rmfamily e.g.,} dark features within UDs) 
which may be close to, if not overlap with the 
telescope's diffraction limit, our photospheric umbral 
time series was created purely from consecutive $4170${\,}{\AA} 
continuum images. These 
images were multiplied by the umbral binary map, producing a 
dataset which could be analysed with Fourier techniques. 
Following the methodology of \citet{Jes07, Jes07b}, the wavelet 
analysis routines of \citet{Tor98} were applied to the $4170${\,}{\AA} 
continuum umbral time series to search for the presence of 
oscillatory behaviour. Considerable oscillatory power was 
present throughout the sunspot umbra, with dominant peaks 
in the Fourier spectrum at $3$ and $5$ minutes. This 
is consistent with the generalised $p$-mode spectrum of 
sunspot oscillations \citep{Lit82, Lit84, Lit86, Bog06}. However, when locations 
of high oscillatory power were examined, it became evident that 
the UDs displayed significant enhancements of wave amplitude, 
at periodicities of approximately $3$~minutes. The 
middle panel of Figure~\ref{fig3} shows the spatial locations where 
oscillatory power at approximately $3$ minutes 
is in excess of $1000$ times the background umbral 
value. To test whether these long-lived umbral brightenings 
oscillate in phase with the surrounding umbra, temporally filtered 
time series of these structures were first created by passing each 
spatially-averaged UD intensity through a 
$2$ -- $4$~minute bandpass filter. The resulting lightcurves, 
dominated by power corresponding to the $3$~minute $p$-mode 
oscillations, are displayed in the upper panel of Figure~\ref{fig4}, 
along with a filtered time series representing a region of the 
quiet umbra devoid of any transient brightenings. It 
is clear that not only do the peaks and troughs of the UD 
oscillations appear similar in time, but these signatures 
also closely follow the oscillations originating from within the 
background umbra.

To quantify any small-scale differences in the oscillating UD 
time series, a phase difference analysis with respect to the 
background umbra was performed. The lower panel of 
Figure~\ref{fig4} reveals the phase difference between each of the 
four UD time series, and an isolated region of the background 
umbra ({\rmfamily i.e.} the solid black line in the upper panel of Figure~\ref{fig4}). 
It is clear that a preference exists for the UDs to oscillate 
in phase with the background umbra. Maximum deviations in the 
phase angle reach approximately $\pm$$40$$\degr$, although these may 
be due to slight drifts in oscillation period between the respective 
time series. For example, the phase angle related to UD1 
(red line in the lower panel of Figure~\ref{fig4}) drifts a total of 
$\approx$$80$$\degr$ over a $700$~s time duration, providing an 
average shift of $\approx$$0.11$$\degr${\,}s$^{-1}$. This drift 
can be explained by a period discrepancy of only $20$~s between 
the UD and the background umbra, something 
which is well within the normal range of $p$-mode frequencies 
\citep{Tho85}. Thus, we interpret the umbra as a single 
oscillating ``drum skin'', which not only induces wave motion in 
its inherent UDs, but also causes these structures to 
oscillate in phase with one another. 

While \citet{Cho09} and \citet{Sta11} have shown that 
$3$~minute magneto-acoustic power is suppressed in sunspot 
umbrae as a result of local absorption and emissivity reduction, 
their spatial resolution was insufficient to allow studies of the 
smallest sunspot features ({\rmfamily e.g.,} umbral dots). Recently, 
\citet{She09} have shown that on small spatial 
scales, the curvature and strength of magnetic field lines 
have substantial influence on the efficiency of magneto-acoustic 
wave propagation. Indeed, these authors suggest that wave 
power can actually be amplified under certain atmospheric 
conditions. Following on from this, our observational data 
suggests that UDs are able to enhance the background 
$p$-mode power by at least three orders-of-magnitude 
(see {\rmfamily e.g.,} middle panel of Figure~\ref{fig3}). 
This may be a result of the magnetic 
field lines acting as efficient conduits for magneto-acoustic 
wave propagation \citep[{\rmfamily e.g.,}][]{Sin92, She06, Kho08, Erd10}.

\newpage
\subsection{Chromosphere}
Chromospheric information comes from the H$\alpha$ core 
imaging dataset, in addition to the Ca~{\sc{ii}} $8542${\,}{\AA} 
spectral imaging scans taken with IBIS. Furthermore, Doppler 
shifts of the Ca~{\sc{ii}} profile minimum allow a series 
of two-dimensional velocity maps to be generated 
\citep{Jes10a}. The Ca~{\sc{ii}} core image shown in the 
right panel of Figure~\ref{fig1} is a true intensity map, 
created by establishing the line-profile minimum at each 
pixel. By displaying Doppler-compensated line-centre intensities, 
rather than rest-wavelength intensities, brightness variations 
throughout the image are more indicative of the source function than of the 
velocities present in the line-forming region \citep{Lee10}.

In a process identical to that applied to the photospheric 
image sequence, the H$\alpha$ and Doppler-compensated Ca~{\sc{ii}} 
data sets were multiplied by the umbral binary map, and subsequently analysed 
using wavelet techniques. As for the $4170${\,}{\AA} continuum 
time series, considerable oscillatory power was 
present throughout the entire sunspot umbra, with dominant peaks 
in the Fourier spectrum at $3$ and $5$ minutes. However, 
remaining consistent with \citet{Tho85}, chromospheric power at 
periodicities of $\sim$$5$~minutes was much reduced when compared 
to those found in the photospheric umbra. To investigate whether 
oscillatory power detected in the chromosphere can be related to similar 
phenomena found in the photosphere, the locations of high oscillatory 
power were examined. The right panel of Figure~\ref{fig3} 
outlines the spatial locations where power, in both intensity and velocity 
signals at approximately $3$ minutes, is in excess of $1000$ times the 
background umbral value. Similar to the locations of high photospheric 
power displayed in the middle panel of Figure~\ref{fig3}, several distinct 
groups manifest in a horseshoe shape in the south-west quadrant of the 
sunspot umbra. 

The locations of significant chromospheric power have 
two distinct differences when compared to their photospheric 
counterparts. First, their spatial positions are slightly offset from those 
displayed in the middle panel of Figure~\ref{fig3}. Secondly, the spatial sizes 
of high chromospheric power are substantially larger than those found in the 
photosphere. These effects can be explained by the geometry of the 
magnetic field lines which stretch outwards from the solar surface. For 
example, high photospheric power encompassing UD2 (middle panel of 
Figure~\ref{fig3}) has a local maximum at the heliocentric co-ordinate 
($-133{\,}.{\!\!}{\arcsec}7$, $-483{\,}.{\!\!}{\arcsec}7$), while the chromospheric 
local maximum is at ($-133{\,}.{\!\!}{\arcsec}1$, $-482{\,}.{\!\!}{\arcsec}2$). Thus, 
the offset between these two maxima is $\approx$$1{\,}.{\!\!}{\arcsec}6$, or 
$\approx$$1100$~km. A separation between the 
$4170${\,}{\AA} continuum and the H$\alpha$/Ca~{\sc{ii}} core formation 
heights of $\sim$$1800$~km \citep{Ver81, Jes12} requires an 
inclination angle of $\approx$$40{\degr}$ to support the assumption 
that the oscillatory signals are from a continuation of the same solar 
structure \citep[{\rmfamily e.g.,}][]{Cen06}. This is the largest offset present 
in our observations, and forms our upper limit of the magnetic field 
inclination angle.
\citet{Rem09} have recently shown that the inclination angles 
(to the vertical) of magnetic flux tubes within sunspot umbrae can be quite large 
($>$$45{\degr}$), with these structures often becoming horizontal 
near the penumbral boundary. Furthermore, \citet{Mar09} utilized 
stereoscopic observations of propagating slow-mode waves in 
coronal structures to infer an absolute inclination 
angle $\sim$$40{\degr}$. Thus, an inclination angle of 
$<$$40{\degr}$ supports the interpretation that oscillatory 
behaviour detected at photospheric and chromospheric layers 
are directly related by the magnetic field lines which extend upwards from the 
solar surface. 

\begin{figure}
\includegraphics[width=\columnwidth]{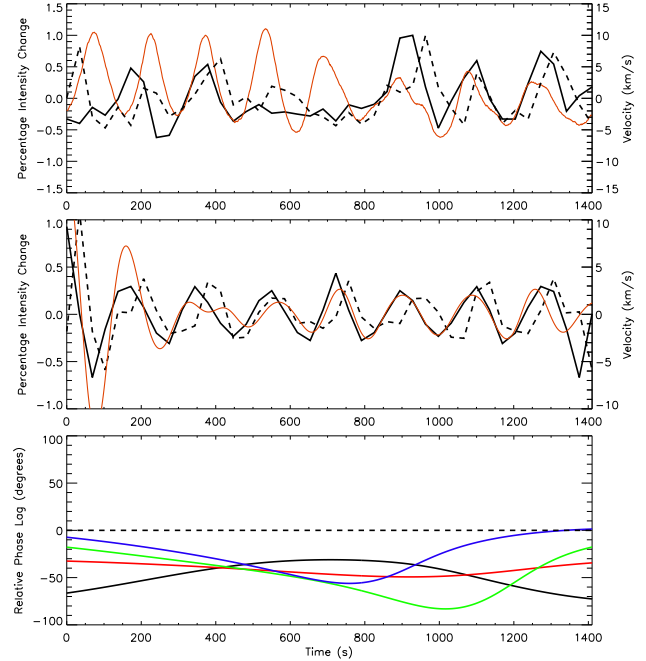}
\caption{{\it{Upper:}} An unfiltered time series, created by spatially averaging 
Ca~{\sc{ii}} intensity (solid line) and velocity (dashed line) signals, along with 
H$\alpha$ core intensities (red line), originating from within 
the contours outlining `UD2' in the right panel of Figure~\ref{fig3}. 
{\it{Middle:}} The same time series from the upper panel, following 
temporal filtering through a $2$ -- $4$~minute 
bandpass filter. A $-90\degr$ phase delay between velocity and 
intensity signals indicates the presence of oscillations which are 
magneto-acoustic in nature, while a close agreement between H$\alpha$ 
and Ca~{\sc{ii}} intensities is a result of their similar formation heights.
{\it{Lower:}} The relative phase angles between the photospheric 
and chromospheric umbral dot oscillations, plotted as a function of time. 
The black, red, green, and blue lines represent umbral dots 
UD1, UD2, UD3, and UD4, respectively, while a black dashed 
line highlights a phase angle of $0$$\degr$.
}
\label{fig5}
\end{figure}

An increase in the area of the oscillating 
regions can be associated with a physical expansion of the magnetic 
flux tubes as a function of atmospheric height. Through examination of 
intense magnetic field concentrations ($>$$1000$~G), \citet{Jes09} were 
able to show that the radial expansion of magnetic flux tubes between 
photospheric and chromospheric 
heights can be as large as a factor of $2$. Assuming these magnetic flux tubes 
demonstrate an expanding cylindrical geometry, doubling the radial 
dimension will result in an area increase of $400$\%. Continuing with the 
previous example, the 
region of high oscillatory power encompassing UD2 covers 
$326$~pixels ($0.8$~Mm$^{2}$ at $2500$~km$^{2}${\,}pixel$^{-1}$) and 
$256$~pixels ($2.5$~Mm$^{2}$ at $10{\,}000$~km$^{2}${\,}pixel$^{-1}$) at 
photospheric and chromospheric heights, respectively. As a result, 
an expansion of only $310$\% is observed, meaning our observations are 
well within the limits of previous expanding magnetic flux tube models 
\citep[{\rmfamily e.g.,}][]{Mei07, Rud08, She10, Kar11, Fed11}.  

The specific mode of oscillation can be determined through investigation 
of the coupling between intensity and velocity signals. Here, we adopt the 
$V-I$ convention \citep{Deu89, Fle89}, which shows the delay of 
maximum intensity ($I$) with respect to maximum blueshift velocity ($V$). 
Thus, a wave which has a velocity signal trailing its intensity signal 
by $1/4$ of a period will have a $V-I$ phase angle of $-90\degr$. 
Due to the Ca~{\sc{ii}} absorption line being sensitive to temperature 
fluctuations \citep{Bee69}, we follow the common practice to adopt the intensity ($I$) as 
a proxy for the local temperature. \citet{Mei77} has shown that when waves 
propagate along moderately inclined 
flux tubes, the phase lag between fluctuations in velocity and temperature 
are the same as those found in purely acoustic waves which have been modified 
by gravity. Therefore, in a fully adiabatic scenario, theory predicts a $V-I$ 
phase angle approaching $0\degr$ for running acoustic waves, 
which can increase to $-90\degr$ when aspects of wave reflection create 
standing acoustic modes \citep{Hof96, Al98, Nig99}. Under isothermal conditions, 
the phase lag can further increase up to $-180\degr$ 
\citep{Mih81, Mih82}. Examination of the upper panel in 
Figure~\ref{fig5} reveals how the unfiltered velocity time series 
(dashed line) trails the co-spatial intensity lightcurve (solid line) by 
approximately $1/4$ of a wave period. This effect becomes 
even more pronounced when the same time series are passed through a  
$2$ -- $4$~minute bandpass filter (middle panel of 
Figure~\ref{fig5}). Chromospheric waves detected within the contours of the 
right panel of Figure~\ref{fig3} have a spatially and temporally averaged 
phase angle of $-87\pm8\degr$, suggesting these waves are magneto-acoustic in 
nature, with characteristics consistent with standing acoustic modes 
\citep{Deu74, Cra78}. The generation of a chromospheric standing 
wave may be the result of a portion of the acoustic wave energy being 
reflected back at the transition region boundary \citep{Sch92, Nak04, Fed11}.

The spatial and temporal averaging of velocity signals found 
in the locations of high chromospheric power results in a net blueshift velocity of 
$\approx$$1.5$~km{\,}s$^{-1}$ (upper panel of Figure~\ref{fig5}). 
While a $V-I$ phase angle of $-87\pm8\degr$ 
implies the presence of wave reflection, a net blueshift velocity may also indicate 
that a portion of the magneto-acoustic wave is propagating in the 
upward direction. To examine the propagation characteristics of the magneto-acoustic 
waves associated with the UDs, spatially-averaged H${\alpha}$ intensity time-series 
were constructed 
for each of the regions contoured in the right panel of Figure~{\ref{fig3}}, and subsequently 
passed through a $2$ -- $4$~minute bandpass filter to isolate oscillatory phenomena 
around 3~minutes. A phase difference analysis was performed between these chromospheric 
H${\alpha}$ lightcurves and their corresponding photospheric $4170${\,}{\AA} continuum 
counterparts created in Section~{\ref{photosphere}}, and displayed in the upper panel 
of Figure~{\ref{fig4}}. A preference for negative phase angles is shown in the lower panel 
of Figure~{\ref{fig5}}, which indicates the oscillatory signals are first observed in the 
photospheric $4170${\,}{\AA} continuum, before later being detected in the chromospheric 
H${\alpha}$ time-series. By averaging the derived phase angles over all UDs, an 
average phase lag of $-43\pm12\degr$ is found, implying the presence of upwardly 
propagating waves. Using the dominant periodicity of $\approx$$3$~minutes, a 
relative phase angle of $-43\degr$ equates to a time lag of 
$\approx$$22$~s. This is consistent with 
the photosphere-to-chromosphere time-lag measurements 
detailed in \citet{Kob11}. However, due to a height 
separation of $\sim$$1800$~km between the $4170${\,}{\AA} continuum and 
H$\alpha$ core image sequences \citep{Ver81, Jes12}, a time lag 
of $\approx$$22$~s requires a phase speed exceeding 
$80$~km{\,}s$^{-1}$. This is clearly unfeasible since a velocity this large will be 
supersonic, and violates our interpretation that the observed waves are 
magneto-acoustic slow modes. However, because the wave trains exist at the 
start of our observing sequence, and continue to be observed as the 
time-series finishes, a factor of $n2\pi$ ($n360\degr$), where $n$ is an 
integer, may be absent from our derived phase angle. By considering 
$n=1$, the phase lag increases to $\approx$$-403\degr$, with the resulting 
phase speed reducing considerably to 
$\approx$$8$~km{\,}s$^{-1}$. Only by observing the start and/or end of a 
propagating wave train can the exact time lag (and associated phase speed) 
be conclusively determined. Nevertheless, the detected waves are 
best described as upwardly-propagating magneto-acoustic modes, which 
travel along magnetic flux tubes inclined to the vertical by 
less than $\approx$$40\degr$. Such an inclination angle may explain why 
\citet{Kob08, Kob11} were unable to correlate (on a pixel-by-pixel basis) 
high chromospheric oscillatory power to that occurring in the underlying 
photosphere. 

Our H$\alpha$ dataset has a much higher cadence than the 
Ca~{\sc{ii}} core image sequence ($1.26$~s instead of $43.4$~s). 
Thus, examination of the H$\alpha$ intensity time series can substantially 
reduce the associated errors when deriving the periodicities of the propagating 
magneto-acoustic waves. As can be seen in 
Figure~\ref{fig5}, the H$\alpha$ data displays prominent intensity oscillations, 
most of which last for the entire duration of the time series. A spatial and 
temporal averaging over the regions encompassed by the red contours 
in the right panel of Figure~\ref{fig3} yields a peak chromospheric 
periodicity of $168\pm7$~s, consistent with the generalised $p$-mode 
spectrum of chromospheric sunspot 
oscillations \citep{OSh01, OSh02, Ban02}. The close resemblance 
between H$\alpha$ and Ca~{\sc{ii}} intensity time series suggests their 
formation heights are remarkably similar. This is also apparent by the 
similarities present in simultaneous snapshots through their respective 
filters ({\rmfamily e.g.,} Figure~\ref{fig1}).

\subsection{Transition Region and Corona}

Due to the reduced spatial resolution of AIA images 
($2$~pixel resolution $\approx$$1{\,}.{\!\!}{\arcsec}2$ $\approx$$870$~km), 
when compared with our simultaneous datasets of the lower solar 
atmosphere ($2$~pixel resolution $\approx$$0{\,}.{\!\!}{\arcsec}138$ $\approx$$100$~km), 
it is imperative to concentrate on the larger-scale structures which 
will be apparent in all image sequences. Following inspection of the 
co-aligned images displayed in the lower panels of 
Figure~\ref{fig2}, it is clear that a number of fan 
structures rise out of the photospheric sunspot umbra, 
and into the corona. These structures are particularly visible 
as intensity enhancements in the coronal AIA images, 
where the $131${\,}{\AA}, $171${\,}{\AA}, $193${\,}{\AA}, 
$211${\,}{\AA}, and $335${\,}{\AA} bandpasses show similar structuring 
extending outwards in the south-west direction. 
However, these fan structures are mostly absent in the 
transition region $304${\,}{\AA} images.

\begin{figure*}
\includegraphics[width=\textwidth]{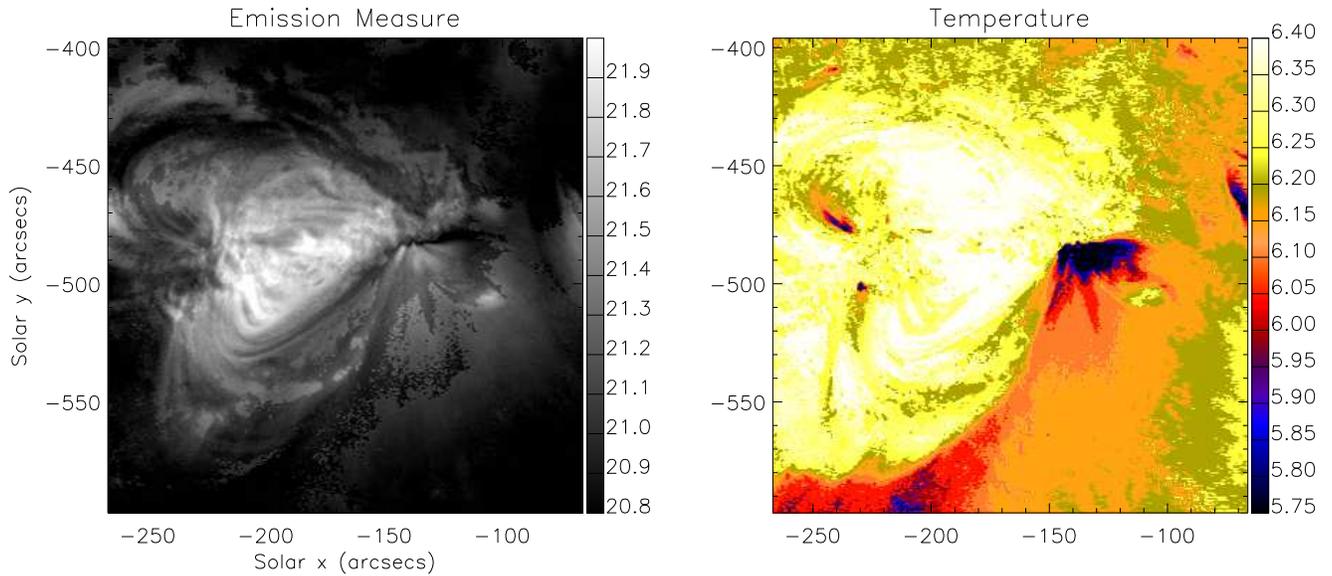}
\caption{Two-dimensional emission measure (in units of 
cm$^{-5}${\,}K$^{-1}$; left) and temperature maps (in units of 
log($T_{e}$); right), derived from near-simultaneous EUV snapshots 
of NOAA~AR~$11250$. The coronal structures currently under 
investigation are located towards the south-west of the active region, 
where temperatures in the range of $0.6$ -- $1.2$~MK dominate.
}
\label{fig6}
\end{figure*}

To investigate whether an absence of the fan 
brightenings in the $304${\,}{\AA} images is a consequence 
of the structures lying outside of the filter's 
temperature response curve, differential emission measure 
(DEM) techniques were employed \citep[{\rmfamily e.g.,}][]{Han12}. 
Utilising the six coronal 
EUV channels on AIA, we were able to construct 
emission measure ($EM$) and temperature ($T_{e}$) estimates 
of active region NOAA~$11250$, including its immediate vicinity. 
Following the methodology presented by \citet{Asc11}, each batch 
of near-simultaneous EUV exposures allowed us to 
construct a time series of $EM$ and $T_{e}$ variables, 
thus allowing the evolutionary changes in each sequence to 
be studied. Importantly, the temperature of the coronal fan 
structures currently under investigation is in the range of 
$0.5$ -- $1.2$~MK (right panel of Figure~\ref{fig6}). Their relatively 
cool temperature, when compared to the multi-million degree 
values found at the centre of the active region, probably 
manifest as a result of open magnetic field lines which cannot 
trap heated plasma. An alternative explanation could revolve around 
their connection with quasi-separatrix layers, which may be 
subject to a peculiar heating function \citep{Sch10}. 
While the $304${\,}{\AA} bandpass has two distinct temperature 
response functions covering approximately $50{\,}000$~K and $1.5$~MK, 
the fact that active region NOAA~$11250$ was positioned close 
to solar disk centre suggests that the resulting images should be 
dominated by the He~{\sc{ii}} emission formed at 
$\sim$$50{\,}000$~K, with contributions from the 
$1.5$~MK Si~{\sc{xi}} $202.22${\,}{\AA} emission line 
minimal \citep{Odw10}. This helps to explain 
why features near one million degrees ({\rmfamily e.g.,} the 
fan structures extending outwards from the underlying sunspot), 
are not readily apparent in the transition region data. 

\begin{figure}
\includegraphics[width=\columnwidth]{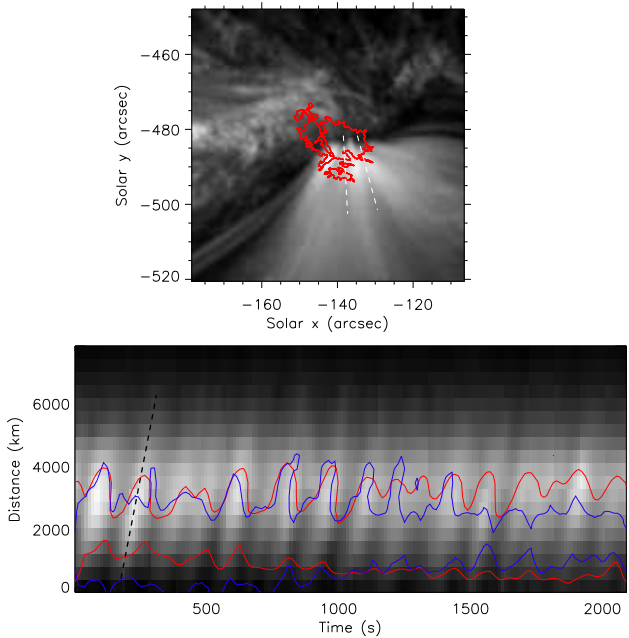}
\caption{An AIA $171${\,}{\AA} image (top), co-spatial with our 
field-of-view of the lower solar atmosphere. Dashed white 
lines outline a coronal fan structure, originating from within the underlying 
sunspot umbra, where propagating wave phenomena are readily apparent. 
The perimeter of the sunspot umbra is highlighted by a solid red line, 
demonstrating how the coronal fan is anchored into regions of high 
chromospheric oscillatory power (see e.g., the right panel in Figure~\ref{fig3}). A 
sample time-distance diagram of this fan is displayed in the lower panel, 
where $0$~km represents the sunspot umbra. Red and blue 
contours outline $211${\,}{\AA} and $335${\,}{\AA} intensities, respectively, 
which are $60${\%} above the local quiescent background. A black dashed line 
represents the propagation of a typical wave front, where the gradient 
provides the wave speed, typically $45\pm7$~km{\,}s$^{-1}$. Wave phenomena 
which are co-spatial, co-temporal, and propagating with the same wave speed are 
readily apparent over a range of coronal temperatures.}
\label{fig7}
\end{figure}

Examination of time-lapse movies of coronal EUV images 
revealed clear and distinctly periodic outflows along the 
coronal fans. To quantify the associated 
periodicities and flow velocities, a series of one-dimensional
slits were placed along the motion path in each of the 
coronal EUV channels. The resulting time-distance cuts 
reveal numerous propagating wavefronts, indicated by 
straight, diagonal trends in the bottom panel of 
Figure~\ref{fig7}. The AIA 
EUV images, in the temperature range of 
$0.4$ -- $2.8$~MK (incorporating the 131{\,}{\AA}, 171{\,}{\AA}, 
193{\,}{\AA}, 211{\,}{\AA}, and 335{\,}{\AA} bandpasses), 
display a dominant periodicity of 
$172\pm17$~s, and a propagation velocity of
$45\pm7$~km{\,}s$^{-1}$. Red and blue contours 
in the lower panel of Figure~\ref{fig7} 
outline the intensity signals present in the higher temperature 
$211${\,}{\AA} and $335${\,}{\AA} bandpasses, respectively. It 
is clear that not only do the periodicities of the wave fronts 
closely resemble one another in both space and time, but 
the intensity gradients (and hence wave speeds) are 
also similar. Propagating wave 
fronts are observed co-spatially, and simultaneously in 
all but the $94${\,}{\AA} AIA bandpass. The signal-to-noise 
of the $94${\,}{\AA} channel is too low 
to extract a time series of sufficient quality for analysis. This 
may be compounded by the fact that the fan structures 
demonstrate a temperature much lower than the $94${\,}{\AA} 
channel's peak response.

A periodicity of $\approx$$172$~s and a propagation 
velocity $\approx$$45$~km{\,}s$^{-1}$ are 
consistent with \citet{Kin03}, who utilised the $171${\,}{\AA} 
and $195${\,}{\AA} filters onboard the TRACE satellite to 
reveal how the propagation velocity of such waves will 
not exceed the local sound speed, even with a large 
inclination angle of the wave guides away 
from the observer's line-of-sight. The typical sound 
speed associated with the dominant Fe{\,}{\sc{ix}} 
emission from the $171${\,}{\AA} bandpass is 
$\approx$$150$~km{\,}s$^{-1}$ \citep{DeM06}, which 
requires the fans currently under investigation 
to have an inclination angle exceeding $70\degr$ 
before the observed wave motion would become 
supersonic. An inclination angle this large is highly 
unrealistic due to the location of the active region on 
the solar disk, in addition to previous surveys of coronal 
magnetic geometries \citep[{\rmfamily e.g.,}][]{Asc02, Asc08, Asc09}. 
Thus, we can conclude that the observed wave phenomena 
is best described as propagating slow-mode waves.

\begin{figure}
\includegraphics[width=\columnwidth]{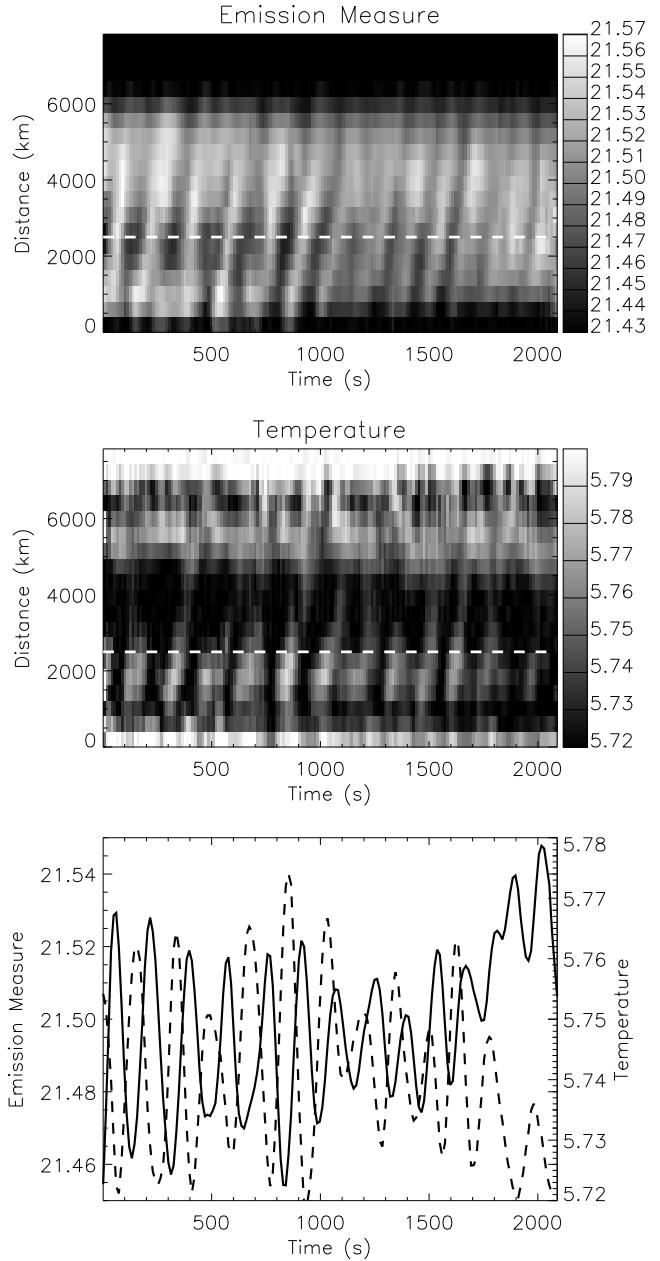}
\caption{Time-distance diagrams of the emission measure (in units of 
cm$^{-5}${\,}K$^{-1}$; top) and temperature (in units of 
log($T_{e}$); middle), derived for the fan structure where 
propagating wave phenomena in EUV intensity images was detected. 
The dashed white lines highlight the spatial position along the slice 
($2500$~km from the underlying umbra) where $EM$ and $T_{e}$ 
time series were created. These values are displayed in 
the lower panel, where the solid line represents the emission measure, 
and the dashed line traces the temperature. Both time series are 
displayed in their native units, as used in the top and middle panels. 
A clear anti-correlation between the $EM$ and $T_{e}$ 
time series is readily apparent.}
\label{fig8}
\end{figure}

Since propagating wave fronts are detected in many of 
the AIA EUV bandpasses, emission measure and 
temperature maps were subsequently investigated for 
the presence of similar signatures. As the emission measure 
maps created here are the sum of (squared) electron densities 
along a given line-of-sight, 
the total mass distribution, regardless of the local temperature, 
can be studied as a function of time \citep{Asc11}. 
Furthermore, the derived pixel-by-pixel temperatures correspond 
to the peak of the DEM distribution for a given line-of-sight, thus 
allowing the entire temperature range of the coronal plasma to be 
easily studied \citep{AscBoe11}. Applying the same 
techniques used on the EUV image sequences, time-distance 
diagrams of the $EM$ and $T_{e}$ variables derived for 
the fan structures reveal identical periodicities and propagation 
velocities as found in the EUV images (middle and upper 
panels of Figure~\ref{fig8}). The 
emission measure is usually defined as \citep{Kin82, Doy85},
\begin{equation}
EM (T_{e}) = \int n_{e}^{2}~dh \ ,
\end{equation}
where $n_{e}$ is the electron density and $h$ is an emitting depth 
along the observer's line-of-sight. Since the compressive phase of a 
magneto-acoustic wave mode will cause the emitting volume 
to decrease, thus reducing the overall emission measure, one would 
expect the $EM$ time series to oscillate in phase with the 
corresponding EUV intensities. However, as the plasma becomes 
compressed, the associated kinetic temperature will increase, 
thus causing the $T_{e}$ signal to oscillate out-of-phase with the 
$EM$ and EUV intensity signals. Indeed, perturbations in the EUV 
intensity, depicted in Figure~\ref{fig7}, are found to occur in phase 
with the derived emission measure fluctuations. Furthermore, 
the lower panel in Figure~\ref{fig8} reveals how oscillations in 
$T_{e}$ are found to be $180\degr$ out-of-phase with respect to 
those detected in the $EM$ signal. Quantitatively, the detected 
fluctuations in the $EM$ and $T_{e}$ values are typically in the 
range $21.50\pm0.03$~cm$^{-5}${\,}K$^{-1}$ and 
$0.55\pm0.03$~MK, respectively. 
This strengthens our interpretation 
that the observed wave phenomena are best described as 
magneto-acoustic slow modes.

\subsection{Numerical Simulations}

\begin{figure*}
\begin{center}
\includegraphics[width=12cm]{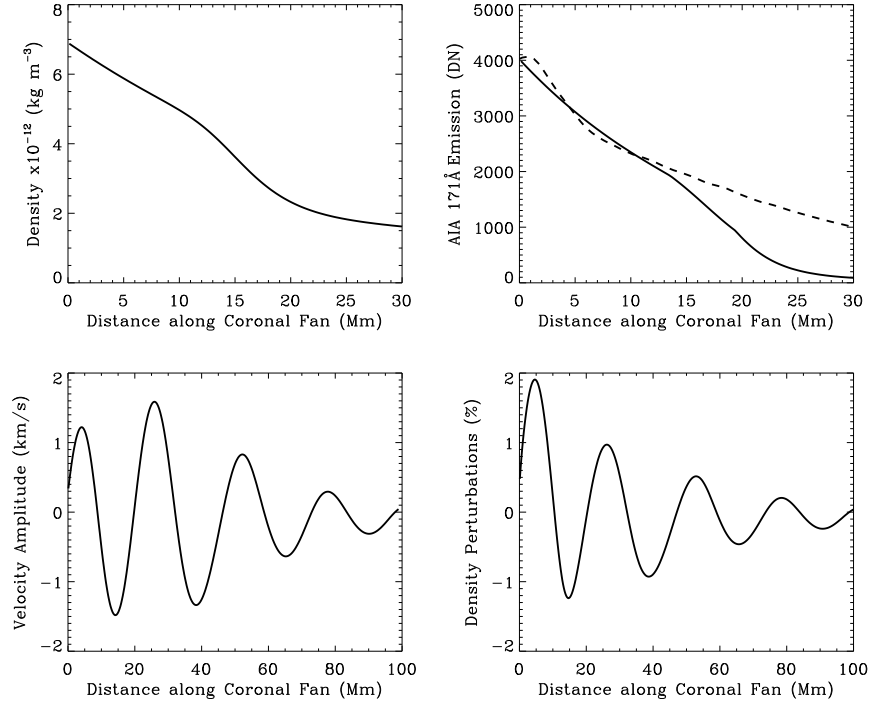}
\caption{The initial (equilibrium) model density (upper-left) used in our 
1D {\sc{lare}} simulations, displayed 
as a function of distance along the coronal fan. The 
forward-modelled AIA 171{\,}{\AA} emission (in DN{\,}pixel$^{-1}$; solid line) 
is plotted in the upper-right panel, alongside the actual observed 
intensity profile (dashed line). Velocity amplitudes, induced by 
driving a slow magneto-acoustic wave into the numerical domain, are 
displayed in the lower-left panel, while the lower-right panel reveals the 
corresponding relative density perturbations, generated through subtraction 
of the forward-modelled values at $t=30$~minutes from those at the 
beginning of the simulation. Damping, caused 
primarily by thermal conduction, is readily apparent in the lower two panels. 
All horizontal axes are in units of Mm, where $1$~Mm $=$ $1000$~km.}
\label{fig9}
\end{center}
\end{figure*}

To model our observational findings, we performed 
a series of 1D numerical simulations using the Lagrangian-Remap 
code \citep[{\sc{lare}};][]{Arb01}, in which thermal conduction and optically thin 
radiation have been included \citep{Owe09}. The $1$D model setup 
is constructed to closely match the observed AIA~$171${\,}{\AA} emission at 
the lowest coronal part of the fan structure. 
In particular, the density profile (upper-left panel of Figure~\ref{fig9}) 
along the fan includes gravitational stratification, 
where the initial ($z=0$) value is determined through equalisation of 
the forward-modelled and observed $171${\,}{\AA} intensities, 
as can be seen in the upper-right panel of Figure~\ref{fig9}. 
Here, the solid line corresponds to the spatial variation in $171${\,}{\AA} emission, 
obtained through forward-modelling of the temperature and density 
profiles inferred from the observed fan structure 
\citep[where the forward modelling is undertaken using a code developed by][]{DeM08}.
A model temperature profile is constructed which utilises the minimum and maximum 
values of the (isothermal) temperatures extracted directly from the 
AIA emission. The background magnetic field is taken to be constant. From 
the upper-right panel of Figure~\ref{fig9}, it is clear that the theoretical model is in 
close agreement with the observed $171${\,}{\AA} emission during the first 
$15$~Mm along the fan. After this, the model emission declines more steeply than 
the observed $171${\,}{\AA} emission. However, we are primarily concerned 
with the behaviour within the first $15$~Mm, and therefore can neglect larger 
distances where the emission becomes faint, and the observations become more 
affected by detector noise. 

A slow magneto-acoustic wave is driven into the domain at the lower boundary, 
through periodic perturbations of the field-aligned velocity component. 
A snapshot of the velocity perturbations at $t=30$ minutes, using a driving 
period of $172$~s to match that of the observations, 
is shown in the lower-left panel of Figure~\ref{fig9}. A slow wave will 
propagate along the field at the characteristic tube speed, which is of the order 
of the local sound speed. Incorporating a temperature $\approx$$1$~MK, and 
a background magnetic field strength $\approx$$10$~G, the tube speed 
is of the order of $122$~km{\,}s$^{-1}$. Combined with a driving period of 
$172$~s, we expect a wavelength of the order of 22~Mm, 
which can be verified in the lower panels of Figure~\ref{fig9}. After an initial increase, 
due to the rapidly decreasing background density, the velocity amplitudes are strongly 
damped through a combination of thermal conduction, optically thin radiation, 
and compressive viscosity. The corresponding density perturbations (i.e., 
the density profile at $t=0$ has been subtracted from the density at 
$t=30$ minutes) are shown in the lower-right panel of Figure~\ref{fig9}. 
Such strong damping is predominantly caused by 
thermal conduction, as was previously uncovered by 
\citet{DeM03, DeM04} and \citet{Owe09}. Only a small fraction of the overall 
damping can be attributed to optically thin radiation and compressive viscosity.

To achieve a more meaningful comparison with the perturbations observed by 
the AIA instrument, we utilised the model density and energy (temperature) 
distributions along the loop, to forward-model the emission for different AIA channels, 
at a multitude of time steps. 
We find that the modelled $171${\,}{\AA} emission is significantly more intense over the first 
$15$~Mm of the coronal fan structure, when compared to the 
$193${\,}{\AA} and $211${\,}{\AA} intensities. Although these AIA channels contain 
some emission lines at cooler temperatures, the $193${\,}{\AA} and $211${\,}{\AA} 
bandpasses mainly respond to hotter coronal temperatures 
\citep[i.e., above $1.5$~MK;][]{Odw10, Bro11}, and hence only begin to display elevated 
intensities when the density reduces and the local temperature increases beyond $1$~MK. 
This is consistent with the AIA observations, whereby the cooler $171${\,}{\AA} 
emission dominates the first $15$~Mm from the underlying sunspot. 
When forward-modelling intensity oscillations, it is important to remember that a variation in 
amplitude of the intensity perturbations is a combination of the change in amplitude of 
the model density and temperature fluctuations, as well as the individual response functions 
of the AIA channels to different temperatures. 

\begin{figure}
\begin{center}
\includegraphics[width=\columnwidth]{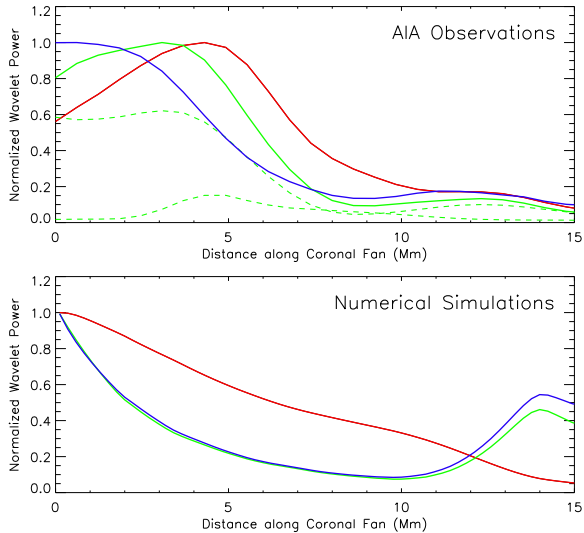}
\caption{Time-averaged global wavelet power spectra, for the 
observed (top) and simulated (bottom) AIA time series, plotted as a 
function of distance along the coronal fan. The red, 
green, and blue lines highlight the oscillatory power contained within 
the $171${\,}{\AA}, $193${\,}{\AA}, and $211${\,}{\AA} bandpasses, 
respectively, with each power spectrum normalised for clarity. Dashed 
green lines displayed in the observed spectra (top) relate to the 
power associated with the hot ($\sim$$1.6$~MK) 
and cool (${\lesssim}$$1$~MK) spectral components 
of the $193${\,}{\AA} channel, with the cool component closely 
following the $171${\,}{\AA} trend, and the hot component tending 
more towards the $211${\,}{\AA} spectrum. The damping of 
magneto-acoustic wave power, at hotter temperatures first, is an 
indication that thermal conduction is the dominant damping mechanism.
}
\label{fig10}
\end{center}
\end{figure}

Finally, we utilise the global wavelet power, at each spatial 
position along the coronal fan, to compare the amplitude decay rates 
found in the different AIA bandpasses in a more quantitative way. To do this, 
a wavelet transform was performed at each position along the fan, with the 
temporally-averaged global wavelet power subsequently calculated. 
The red, green, and blue lines in Figure~\ref{fig10} 
represent the time-averaged oscillatory power around 3~minutes, for the 
AIA $171${\,}{\AA}, $193${\,}{\AA}, and $211${\,}{\AA} observed and 
forward-modelled bandpasses, respectively. Note that each 
of these have been normalised to their individual maxima for display 
purposes. Of particular note is the observed $193${\,}{\AA} power 
spectrum. Here, the solid green line represents the oscillatory power 
registered through the entire AIA bandpass, while the dashed green lines 
relate to the power associated with the hot ($\sim$$1.6$~MK) 
and cool (${\lesssim}$$1$~MK) spectral components 
of this channel. These components have been separated using the 
methods detailed in \citet{Kid12}, with the resulting cool emission peaking 
alongside the $171${\,}{\AA} power at $\approx$$4.5$~Mm, and the 
hot emission displaying a peak between the $171${\,}{\AA} and 
$211${\,}{\AA} bandpasses ($\approx$$3$~Mm). In reality, the global wavelet power of the 
$193${\,}{\AA} and $211${\,}{\AA} emission is substantially 
lower than that of the $171${\,}{\AA} emission. However, the time-averaged power of the 
$193${\,}{\AA} and $211${\,}{\AA} modelled intensities are very similar, as 
is the case for the observations. 
For these wavelengths, the simulated global wavelet power reaches a minimum around 
$10$~Mm, while the $171${\,}{\AA} bandpass reaches its minimum power slightly 
further along the fan structure, at about $15$~Mm. 
These results agree qualitatively with the observed time-averaged 
power, where the $193${\,}{\AA} and $211${\,}{\AA} channels near-simultaneously 
reach a minimum at about $9$~Mm along the coronal fan, with the 
$171${\,}{\AA} power diminishing completely by $15$~Mm. This reiterates the 
importance of thermal conduction in the damping of coronal slow-mode waves. 
The increase in time-averaged global wavelet power, after $\approx$$10$~Mm 
in both the simulated and observed $193${\,}{\AA} and $211${\,}{\AA} time 
series, is due to the fact that, despite the decreasing amplitudes of the 
model density and temperature, the percentage intensity perturbations in 
these channels are actually increasing. This is a result of the increased 
sensitivity of the AIA $193${\,}{\AA} and $211${\,}{\AA} bandpasses 
to the $>$$1$~MK temperatures which have been reached by this 
distance along the fan structure. The increase is not as apparent 
in the observed $193${\,}{\AA} and $211${\,}{\AA} time-averaged wavelet 
power spectra, which may be due to the decreased signal-to-noise levels 
found in these channels, especially when compared to the 
$171${\,}{\AA} bandpass.

\section{Overview and Concluding Remarks}

Here we present high-cadence observations of the solar 
atmosphere, obtained using the latest ground- and space-based 
facilities. Prominent oscillatory behaviour is detected 
throughout the optical and EUV image sequences, with 
remarkable similarities found between the detected wave modes. 
First, a number of UD 
structures in the photospheric umbra are found to display intensity 
oscillations with a 
$\approx$$3$~minute periodicity. These oscillations exhibit 
considerable power, with regions encompassing the UDs displaying 
more than three orders-of-magnitude stronger power than the 
background umbra (Figure~\ref{fig3}). Next, chromospheric intensity 
{\it{and}} velocity 
measurements were analysed for the presence of co-spatial 
and co-temporal oscillations. Such phenomena were detected, both 
on larger spatial scales, and with small central offsets with respect 
to the underlying photospheric oscillations. By considering the 
extension of magnetic flux tubes from the solar surface out into 
the upper solar atmosphere, a geometric expansion of only 
$76$\% in the radial direction and an inclination angle 
$<$$40\degr$ allows the observed oscillations to be interpreted 
as originating from within the same magnetic flux tube.

Following examination of the phase lag between chromospheric velocity 
and intensity components, a $V-I$ phase angle of 
$-87\pm8\degr$ was derived, allowing these waves to be described 
as a magneto-acoustic mode, with characteristics consistent with 
standing acoustic modes (middle panel of Figure~\ref{fig5}). 
The generation of a chromospheric standing 
wave may be the result of partial wave reflection at the 
transition region boundary. An average blueshift velocity of 
$\approx$$1.5$~km{\,}s$^{-1}$ 
was found in the locations where high chromospheric oscillatory 
power was present. This is a sub-sonic velocity, and coupled with a time lag 
between photospheric and chromospheric oscillatory phenomena, 
strengthens our interpretation that the observed oscillations are 
upwardly-propagating magneto-acoustic 
waves, which originate in UD structures located in the photospheric 
umbra. 

A prominent fan structure is present in the simultaneous coronal 
EUV images, namely those from the AIA $131${\,}{\AA}, $171${\,}{\AA}, 
$193${\,}{\AA}, $211${\,}{\AA}, and $335${\,}{\AA} bandpasses. This fan 
is not readily apparent in either the transition-region dominated 
$304${\,}{\AA} emission, or in the higher temperature ($\sim$$7.0$~MK) 
$94${\,}{\AA} bandpass. Using DEM techniques, we 
constrained the temperature of the coronal fan to $0.5$ -- $1.2$~MK, 
thus placing it outside the temperature range of both the 
$94${\,}{\AA} and $304${\,}{\AA} filtergrams (Figure~\ref{fig6}). Time-distance 
techniques were employed on the EUV imaging data where the fan 
structure was readily apparent, allowing the characteristics of propagating 
wave phenomena to be uncovered. Most coronal channels, regardless 
of their absolute temperature sensitivity, revealed outwardly propagating 
wave fronts with an average periodicity and velocity of 
$172\pm17$~s and $45\pm7$~km{\,}s$^{-1}$, respectively 
(lower panel of Figure~\ref{fig7}). The out-of-phase nature between the derived 
temperature and emission measure signals indicates the presence 
of a compressive wave mode (Figure~\ref{fig8}). This, coupled with a sub-sonic 
wave speed ($\approx$$45$~km{\,}s$^{-1}$), highlights the fact 
that these coronal phenomena are best described as upwardly-propagating 
magneto-acoustic slow mode waves. 

Employing numerical simulations, we 
were able to accurately simulate the behaviour of the coronal EUV emission. 
Utilising input parameters derived directly from the AIA observations, 
forward-modelling techniques allowed us to evolve velocity, density, and 
emissivity values forward in time, creating a time series which could be 
directly compared to the AIA observations. Crucially, our 
simulations revealed that thermal conduction is the primary damping 
mechanism behind the dissipation of magneto-acoustic slow-mode waves 
in the corona. Other 
mechanisms, including optically thin radiation and compressive viscosity, 
play a secondary role in the damping of these oscillations.

The fan structure observed in the AIA images, which displays 
signatures of propagating magneto-acoustic waves, 
appears to have anchor points in the 
south-west quadrant of the photospheric sunspot umbra (lower 
panels of Figure~\ref{fig2}). These locations 
are also consistent with the presence of large-amplitude 
wave phenomena detected in simultaneous photospheric and 
chromospheric image sequences. The co-temporal and 
co-spatial relationship between these upwardly-propagating 
magneto-acoustic wave modes, detected throughout the 
entire solar atmosphere, suggests such coronal phenomena 
may be driven by UD oscillations occurring inside the sunspot umbra. 
With this conclusion, it appears that photospheric 
structures which are on the order of $0{\,}.{\!\!}{\arcsec}5$ ($360$~km) in 
diameter, can have a strong influence on coronal structures 
not only several thousand km above their position, but on structures 
which have expanded into the local plasma to diameters often 
exceeding $10${\arcsec} ($7000$~km).

\acknowledgments
DBJ thanks the Science and Technology Facilities Council (STFC) for a Post-Doctoral Fellowship. 
IDM acknowledges support from a Royal Society University Research Fellowship. 
PHK is grateful to the Northern Ireland Department of Education and Learning for a PhD studentship. 
DJC thanks the CSUN College of Science for start-up funding related to this project. 
Solar Physics research at QUB is supported by STFC. 
The ROSA project is supported by The European Office of Aerospace Research \& Development.
HARDcam observations were made possible by a Royal Society Research Grant (2009).
We are grateful for the use of SDO/AIA images, which were obtained courtesy of NASA/SDO and the 
AIA, EVE, and HMI science teams.



{\it Facilities:} \facility{Dunn (ROSA, HARDcam, IBIS)} \& \facility{SDO (AIA)}.

\end{document}